 \documentclass[aps,twocolumn,pra,superscriptaddress,letterpaper,nofootinbib,longbibliography]{revtex4-2}
\usepackage{amssymb}
\usepackage{physics}
\usepackage{mathtools}
\usepackage{upgreek}
\usepackage{booktabs}
\AtBeginDocument{
	\heavyrulewidth=.08em
	\lightrulewidth=.05em
	\cmidrulewidth=.03em
	\belowrulesep=.65ex
	\belowbottomsep=0pt
	\aboverulesep=.4ex
	\abovetopsep=0pt
	\cmidrulesep=\doublerulesep
	\cmidrulekern=.5em
	\defaultaddspace=.5em
}
\usepackage{lipsum}
\usepackage{subfig}
\usepackage{caption}
\usepackage[justification=Justified,format=plain]{caption}

\usepackage{txfonts}
\usepackage[pdftex]{graphicx}
\usepackage[usenames, dvipsnames, svgnames, table]{xcolor}
\usepackage[colorlinks=true, citecolor=Blue, linkcolor=BrickRed, urlcolor=cyan]{hyperref}
\usepackage{bm}
\usepackage{placeins}
\usepackage{setspace}
\usepackage[final]{changes}
\usepackage{circuitikz}

\usepackage{xparse,xcoffins}

\ExplSyntaxOn
\NewCoffin\imagecoffin
\NewCoffin\labelcoffin

\keys_define:nn { miguel/label }
 {
  label   .tl_set:N = \l_miguel_label_tl,
  labelbox .bool_set:N = \l_miguel_label_box_bool,
  labelbox .default:n = true,
  fontsize .tl_set:N = \l_miguel_label_size_tl,
  fontsize .initial:n = \footnotesize,
  pos .choice:,
  pos/nw .code:n = \tl_set:Nn \l_miguel_label_pos_tl { left,up },
  pos/ne .code:n = \tl_set:Nn \l_miguel_label_pos_tl { right,up },
  pos/sw .code:n = \tl_set:Nn \l_miguel_label_pos_tl { left,down },
  pos/se .code:n = \tl_set:Nn \l_miguel_label_pos_tl { right,down },
  pos/n .code:n = \tl_set:Nn \l_miguel_label_pos_tl { hc,up },
  pos/w .code:n = \tl_set:Nn \l_miguel_label_pos_tl { left,vc },
  pos/s .code:n = \tl_set:Nn \l_miguel_label_pos_tl { hc,down },
  pos/e .code:n = \tl_set:Nn \l_miguel_label_pos_tl { right,vc },
  pos .initial:n = nw,
  unknown .code:n   = \clist_put_right:Nx \l_miguel_label_clist
                       { \l_keys_key_tl = \exp_not:n { #1 } }
 }
\clist_new:N \l_miguel_label_clist
\box_new:N \l_miguel_label_box
\box_new:N \l_miguel_label_image_box

\NewDocumentCommand{\xincludegraphics}{O{}m}
 {
  \group_begin:
  \tl_clear:N \l_miguel_label_tl
  \clist_clear:N \l_miguel_label_clist
  \keys_set:nn { miguel/label } { #1 }
  \tl_if_empty:NTF \l_miguel_label_tl
   {
    \miguel_includegraphics:Vn \l_miguel_label_clist { #2 }
   }
   {
    \SetHorizontalCoffin\imagecoffin
     {
      \miguel_includegraphics:Vn \l_miguel_label_clist { #2 }
     }
    \SetHorizontalCoffin\labelcoffin
     {
      \raisebox{\depth}
       {
        \bool_if:NTF \l_miguel_label_box_bool
         { \fcolorbox{white}{white}{\l_miguel_label_size_tl\l_miguel_label_tl} }
         { \l_miguel_label_size_tl\l_miguel_label_tl }
       }
     }
    \SetVerticalPole\imagecoffin{left}{3pt+\CoffinWidth\labelcoffin/2}
    \SetVerticalPole\imagecoffin{right}{\Width-3pt-\CoffinWidth\labelcoffin/2}
    \SetHorizontalPole\imagecoffin{up}{\Height-3pt-\CoffinHeight\labelcoffin/2}
    \SetHorizontalPole\imagecoffin{down}{3pt+\CoffinHeight\labelcoffin/2}
    \use:x{\JoinCoffins\imagecoffin[\l_miguel_label_pos_tl]\labelcoffin[vc,hc]} 
    \TypesetCoffin\imagecoffin
   }
   \group_end:
 }
\NewDocumentCommand{\setlabel}{m}
 {
  \keys_set:nn { miguel/label } { #1 }
 }

\cs_new_protected:Nn \miguel_includegraphics:nn
 {
  \includegraphics[#1]{#2}
 }
\cs_generate_variant:Nn \miguel_includegraphics:nn { V }

\ExplSyntaxOff

\begin{document}

\title{$2\cdot 10^{-13}$ fractional laser frequency stability with a 7-cm unequal-arm Mach-Zehnder interferometer}

%
%

\author{Victor Huarcaya}
\affiliation{Max Planck Institute for Gravitational Physics (Albert Einstein Institute), D-30167 Hannover, Germany\\
Leibniz Universit\"at Hannover, D-30167 Hannover, Germany}

\author{Miguel Dovale \'Alvarez}
\email{miguel.dovale@aei.mpg.de}
\affiliation{Max Planck Institute for Gravitational Physics (Albert Einstein Institute), D-30167 Hannover, Germany\\
Leibniz Universit\"at Hannover, D-30167 Hannover, Germany}

\author{Daniel Penkert}
\affiliation{Max Planck Institute for Gravitational Physics (Albert Einstein Institute), D-30167 Hannover, Germany\\
Leibniz Universit\"at Hannover, D-30167 Hannover, Germany}

\author{Stefano Gozzo}
\affiliation{Max Planck Institute for Gravitational Physics (Albert Einstein Institute), D-30167 Hannover, Germany\\
Leibniz Universit\"at Hannover, D-30167 Hannover, Germany}

\author{Pablo Mart\'inez Cano}
\affiliation{Max Planck Institute for Gravitational Physics (Albert Einstein Institute), D-30167 Hannover, Germany\\
Leibniz Universit\"at Hannover, D-30167 Hannover, Germany}

\author{Kohei Yamamoto}
\affiliation{Max Planck Institute for Gravitational Physics (Albert Einstein Institute), D-30167 Hannover, Germany\\
Leibniz Universit\"at Hannover, D-30167 Hannover, Germany}

\author{Juan Jos\'e Esteban Delgado}
\affiliation{Max Planck Institute for Gravitational Physics (Albert Einstein Institute), D-30167 Hannover, Germany\\
Leibniz Universit\"at Hannover, D-30167 Hannover, Germany}

\author{Moritz Mehmet}
\affiliation{Max Planck Institute for Gravitational Physics (Albert Einstein Institute), D-30167 Hannover, Germany\\
Leibniz Universit\"at Hannover, D-30167 Hannover, Germany}

\author{Karsten Danzmann}
\affiliation{Max Planck Institute for Gravitational Physics (Albert Einstein Institute), D-30167 Hannover, Germany\\
Leibniz Universit\"at Hannover, D-30167 Hannover, Germany}

\author{Gerhard Heinzel}
\affiliation{Max Planck Institute for Gravitational Physics (Albert Einstein Institute), D-30167 Hannover, Germany\\
Leibniz Universit\"at Hannover, D-30167 Hannover, Germany}

\begin{abstract}
To achieve sub-picometer sensitivities in the millihertz band, laser interferometric inertial sensors rely on some form of reduction of the laser frequency noise, typically by locking the laser to a stable frequency reference, such as the narrow-linewidth resonance of an ultra-stable optical cavity or an atomic or molecular transition. In this paper we report on a compact laser frequency stabilization technique based on an unequal-arm Mach-Zehnder interferometer that is sub-nanometer stable at $10\,\upmu$Hz, sub-picometer at $0.5\,$mHz, and reaches a noise floor of $7\,\mathrm{fm}/\!\sqrt{\mathrm{Hz}}$ at 1\,Hz. The interferometer is used in conjunction with a DC servo to stabilize the frequency of a laser down to a fractional instability below $4 \times 10^{-13}$ at averaging times from 0.1 to 100 seconds. The technique offers a wide operating range, does not rely on complex lock acquisition procedures, and can be readily integrated as part of the optical bench in future gravity missions.

\end{abstract}

\maketitle

\section{Introduction}

\begin{figure*}
\centering
\includegraphics[width=\textwidth]{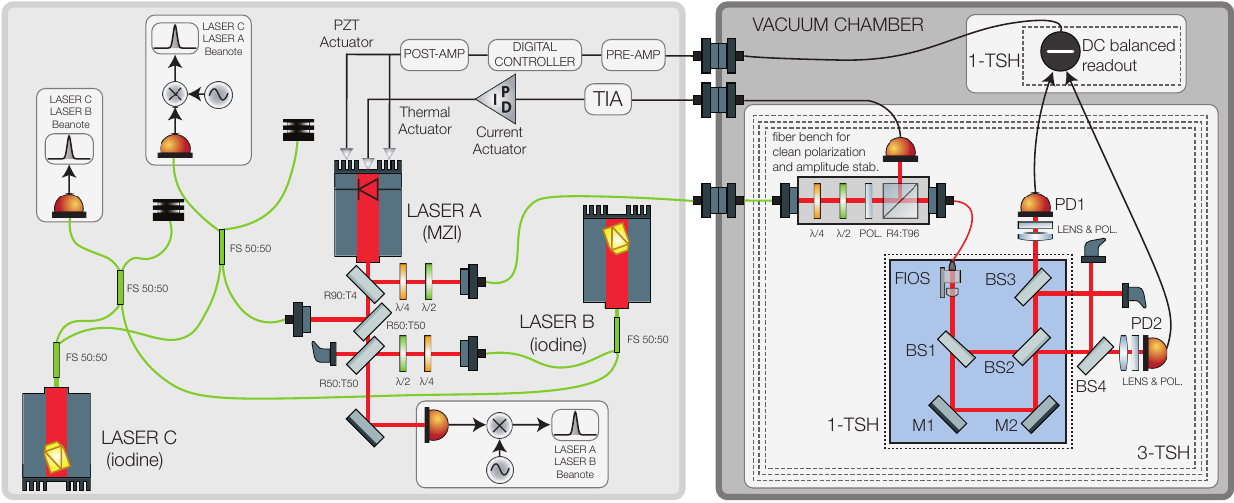}
\caption{Experimental setup. Light from laser A is injected into a vacuum chamber, where it is fed to the ultra-stable interferometer (blue box) via the quasi-monolithic fiber injector optical subassembly (FIOS). The interferometer, along with the auxiliary optics and detectors for polarization cleaning, amplitude stabilization, and interferometric readout, is located inside a high-performance triple-layered thermal enclosure (3-TSH) with an additional single-layer thermal shield (1-TSH) surrounding just the Mach-Zehnder interferometer. The difference-current between the two readout diodes is converted into a voltage by a homemade electronic circuit surrounded by yet another single-layer thermal enclosure (1-TSH). The signal is processed by a digital servo and fed back to the slow and fast frequency actuators of the laser. Laser A is beat with two iodine-stabilized lasers (lasers B and C) to help assess the achieved stability. The beat signals are mixed down to below 100\,MHz with an ultra-stable signal generator and read out via a phasemeter.}
\label{figure:layout}
\end{figure*}

Laser interferometers are a key resource in many areas of science and technology, such as precision metrology, geodesy, and gravitational-wave detection. By measuring tiny distance variations with high precision, interferometers can be used to test fundamental physics~\cite{Mueller2003}, and reveal the gravity field of Earth~\cite{GFO2019, Tapley2019}, or the passing of gravitational waves~\cite{Abbott2016, Abbott2017, Miller2019}.

Laser interferometers aiming to measure displacements with high precision suffer from laser frequency noise coupling proportional to the optical pathlength mismatch $\Delta l$ between the two interfering arms,
\begin{equation}
\frac{\delta l} {\Delta l} = \frac{\delta f}{f_0}
\label{equation:dl-to-df} 
\end{equation}
where $f_0$ is the average frequency of the laser, $\delta f$ is the laser frequency noise, and $\delta l$ is the resulting optical pathlength noise.

For example, the future space-based gravitational wave detector \emph{LISA}~\cite{LPF2015} aims to use laser interferometry to measure picometer-level changes in distance between spacecraft over a baseline of $2.5$-million kilometers, which requires laser frequency noise suppression by many orders of magnitude. Time delay interferometry~\cite{Tinto2003} is proposed to aid the task, setting a target on the laser frequency pre-stabilization of~\cite{LFCST}:
\begin{equation}
\tilde\nu = 300\,\frac{\mathrm{Hz}}{\sqrt{\mathrm{Hz}}} ~\left( \frac{1\,\mathrm{m}}{\Delta L} \right) ~u(f) 
\label{equation:freq-noise}
\end{equation}
for frequencies between 20\,$\upmu$Hz and 1\,Hz, where $\Delta L$ is the absolute ranging accuracy~\cite{Esteban2011, Sutton2013}, currently estimated at 1\,m, and $u(f)$ is the noise shape function given by
\begin{equation}
u(f) = \sqrt{1+\left( \frac{2\,\mathrm{mHz}}{f}\right)^4}
\label{equation:noise-shape}
\end{equation}
which describes a mixture of \emph{white noise} with flat power spectrum, and \emph{random run} noise with $f^{-4}$ power spectrum. 

The current baseline for LISA is using a cavity-stabilized laser similar to the one onboard GRACE Follow-On~\cite{Pierce2012, GFO2019, LRI2019} with a stability roughly an order of magnitude better than Equation~\ref{equation:freq-noise}. Alternative schemes, such as arm-locking~\cite{Sheard2003, Thorpe2005, Sheard2006, Wand2009, Valliyakalayil2022} or stabilization to molecular iodine hyperfine transitions near 532\,nm~\cite{Leonhardt2006, Schuldt2019}, have been proposed. One such scheme involves pre-stabilizing the laser to an unequal-arm Mach-Zehnder interferometer (MZI), similar to the scheme used in the LISA technology package (LTP) interferometer~\cite{Heinzel2003, Heinzel2004} onboard LISA Pathfinder~\cite{LPF2015, LPF2016, LPF2018, LPF2019}. 

Other prominent examples of laser interferometers at the frontier of physics are the gravitational wave detectors \emph{Advanced LIGO}~\cite{Collaboration2015} and \emph{Advanced Virgo}~\cite{AdvancedVirgo15}, which are sensitive to displacements in the 20\,Hz to 5\,kHz band, reaching a noise floor of $2 \cdot 10^{-20}\,\mathrm{m}/\!\sqrt{\mathrm{Hz}}$ at 100\,Hz, ten orders of magnitude below the level of ground motion at the site~\cite{Martynov2016, Buikema2020}. To achieve this amazing stability, seismic noise is reduced by a combination of passive and active stabilization stages~\cite{Carbone2012}. The current schemes are limited by the readout noise of the sensors and lead to excess controls noise below 30\,Hz~\cite{Cahillane2022}. New seismic isolation schemes~\cite{MowLowry2019, Ubhi2021, vanDongen2023} based on laser interferometric readout (see, e.g., References~\cite{Cooper2018, Isleif2019, Yang2020, Zhang2022, Zhang2022a, Kranzhoff2022, Smetana2022, Yan2022}) promise to break the \emph{seismic wall} and lead to sensitivity improvements of current and future detectors at the lower frequencies, with substantial rewards in astrophysical applications~\cite{Yu2018}.

If not addressed, laser frequency noise is one of the leading sources of noise in laser interferometric inertial sensors, particularly below 1\,Hz, even if the macroscopic interferometer arm lengths are matched using best efforts, and even when using commercial narrow-linewidth lasers (e.g., 1\,kHz linewidth for 0.1\,s averaging time). To achieve sub-picometer sensitivities in the millihertz band, some form of reduction of the laser frequency noise is required. The usual schemes involve stabilizing the laser to an ultra-stable optical cavity or an atomic or molecular reference. Such schemes are also commercially available, but they are bulky, costly, and rely on complex electronics.

In a previous article~\cite{Gerberding2017}, a compact quasi-monolithic MZI with an intentional arm length difference of 7\,cm and a DC readout scheme was introduced as a simpler alternative to conventional laser locking schemes. In comparison to an optical cavity or an atomic or molecular reference, the MZI technique offers a wide operating range and does not require a complex lock acquisition procedure. Continuous frequency tuning is possible by purely electronic means and does not require physically changing the resonance frequency of the frequency reference. The MZI in~\cite{Gerberding2017} was shown to provide an impressive long-term dimensional stability, beating the sub-picometer mark at 5\,mHz. For comparison, the LTP interferometer beats this mark at 10\,mHz~\cite{Armano2021}. 

In this paper, we present the next generation of this device, capable of reaching sub-picometer sensitivity at 0.5\,mHz. The combination of an ultra-stable quasi-monolithic fiber injector and a high-performance heat shield system allows us to realize a new benchmark of laser frequency stability with a compact interferometer, yielding a sensitivity improvement of one to nearly two orders of magnitude at low frequencies compared to the previous realization.

Employed in conjunction with an inertial sensor of matching stability, this method enables displacement sensing with a sensitivity better than $1\,\mathrm{pm}/\!\sqrt{\mathrm{Hz}} \cdot u(f)$ down to 10\,$\upmu$Hz, which makes it a promising candidate for both ground-based seismic isolation systems or future gravity missions in space.

\section{Experimental setup}

The experimental setup is depicted in Figure~\ref{figure:layout}. A Mach-Zehnder interferometer with an arm length difference of $\Delta l \approx 7$\,cm is used as an ultra-stable length reference for laser frequency stabilization of laser A, which is a 1064\,nm non-planar ring oscillator (NPRO) laser.

The design and construction of the MZI is described in~\cite{Gerberding2017}, and a summary is given here. The MZI consists of a baseplate made of CLEARCERAM\texttrademark~CCX-HS ultra-stable glass ceramic to which fused silica components are attached via UV adhesive bonding. The baseplate material is chosen for its close to zero coefficient of thermal expansion around room temperature, while fused silica is the material of choice for the optical elements due to its excellent transmission properties. The baseplate has dimensions of $13.5 \times 13.5 \times 3.6 \,\mathrm{cm}^{3}$.

The interferometer design was aided by the C++ optical modeling library IFOCAD~\cite{Ifocad}. In~\cite{Gerberding2017}, an off-the-shelf commercial fiber coupler was used to inject laser light into the West port of the input beam splitter (BS1). In order to improve on the previously reported stability, especially at very long measurement times, a quasi-monolithic fiber injector optical subassembly (FIOS) developed in-house was retrofitted to the North port of BS1. The beam delivered by the FIOS is split into the short and long arms at BS1. The beam traveling along the long arm is reflected off mirrors M1 and M2 before being interfered with the short arm beam at the recombination beam splitter BS2. A third and fourth beam splitters (BS3 and BS4) are placed in the North and East output ports of the beam combiner to allow performing diagnostic measurements, such as optical zero measurements, whilst maintaining symmetry between the photodetectors.

 The precise positioning and orientation of components on the baseplate are optimized via simulations to reduce the impact of spurious beams caused by residual reflections at the secondary surfaces, which have been identified as a critical source of noise in high-precision interferometers~\cite{Katha}. The input beam splitter and the beam combiner are wedged in order to separate the secondary reflections from the main optical path.

\begin{figure}
\centering
\xincludegraphics[width=0.45\textwidth, pos=nw, label=a), fontsize=\large]{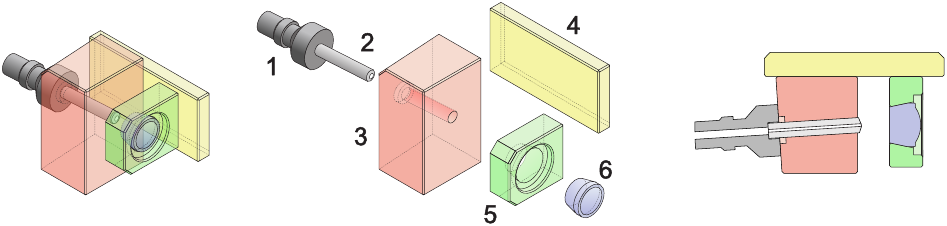}
\xincludegraphics[width=0.45\textwidth, pos=nw, label=b), labelbox, fontsize=\large]{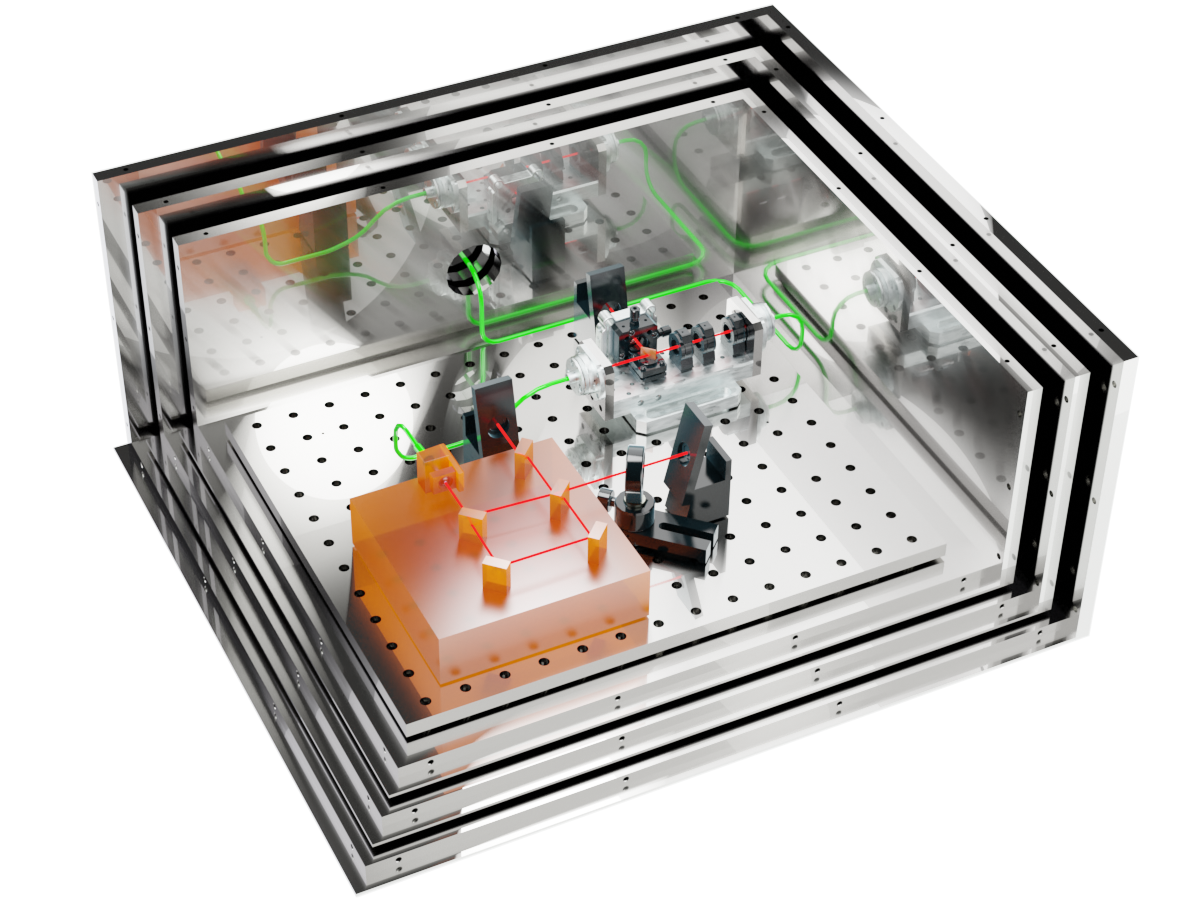}
\caption{The fiber injector optical subassembly (a) consists of a polarization-maintaining 1064\,nm single-mode optical fiber (1) equipped with a bare ferrule (2) held in place by a tightly fitted hole in the fused silica fiber mount (3). An off-the-shelf anti-reflectively coated aspherical lens of D-ZLAF52LA glass, modified for a tapered outer surface (6), is UV-glued into a matching hole in the fused silica lens holder (5). Finally, the fiber mount and the lens holder are joined together at the desired distance, position, and orientation using a longitudinal girder of fused silica (4) and two thermally compensating layers of UV adhesive. The next-generation unequal-arm Mach-Zehnder interferometer (b) consists of an ultra-stable glass ceramic baseplate to which the fused silica components and the fiber injector optical subassembly are bonded via UV adhesive. A set of three aluminum heat shields isolate the interferometer from external temperature fluctuations. The aluminum plate surfaces are polished to lower their emissivity and slow down radiative heat transfer inside the enclosure.}
\label{figure:MZI}
\end{figure}

The FIOS (Figure~\ref{figure:MZI}a) is made of six parts combining a fiber end and a lens into a quasi-monolithic, non-adjustable package, thus significantly reducing the effects of both mechanical and thermal creep in comparison to conventional fiber injectors. The FIOS was first pre-assembled and then installed on the MZI with all other components already fixed to the baseplate. The alignment of the FIOS was done with the help of a homemade positioning device and continuous contrast monitoring by applying a deep frequency modulation~\cite{Gerberding:15, Isleif:16} to the laser. The final contrast achieved was $94\%$, and no discernible contrast degradation was observed following a two-year operational period encompassing several vacuum cycles, highlighting the satisfactory long-term stability of the UV adhesive bonding.

The detection is performed by two identical 50\,mm$^2$ circular active area silicon PIN photodiodes located at the complementary output ports of BS2. A focusing lens is placed in front of each photodiode to help minimize transverse beam jitter, and we incorporate thin-film polarizers with high extinction ratios mounted directly in front of the photodiodes to mitigate the impact of parasitic interferences arising from residual beams with orthogonal polarizations. 

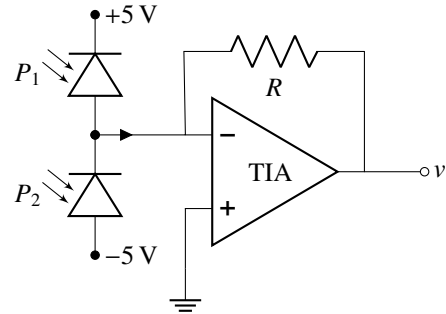
\begin{figure}
\begin{center}
\begin{circuitikz}[scale=0.8]
\draw
(-1.5,1) node[anchor=west]{$P_1$}
(-1.5,-1) node[anchor=west]{$P_2$}
(0,0) to[empty photodiode, *-*] (0,2)
(0,-2) to[empty photodiode, *-*] (0,0)
(0,0) to[short, -, i=$~$] (1,0)
(1,0) to[short, -] ++(0.5,0) coordinate(FB)
node[op amp, anchor=-](OA){TIA}
(OA.+) -- ++(0,-1) 
to[short, -] ++(0,0) node[ground]{}
(OA.out) to[short, -o] ++(1,0) node[right]{$v$}
(OA.out) -- ++(0,2) coordinate(FBR)
(FBR)to[R=$R$, -] ++(-3,0)
-- (OA.-)
(0,2) node[anchor=west] {$+5$\,V}
(0,-2) node[anchor=west] {$-5$\,V};
\end{circuitikz}
\end{center}
\caption{Schematic of the balanced differential trans-impedance amplifier (TIA).}
\label{figure:TIA}
\end{figure}

The photodiodes are operated in reverse bias voltage and connected in a balanced differential trans-impedance amplifier (TIA) performing a direct current subtraction. The basic schematic of the sensor is depicted in Figure~\ref{figure:TIA}. The power at each photodiode depends on the laser frequency $f$ and is given by
\begin{align}
P_1(f) &= p_1 \left[ 1 + c_1 \cdot \cos \left(\frac{2\pi f \Delta l}{c} + \varphi_0 \right) \right] \nonumber\\
P_2(f) &= p_2 \left[ 1 - c_2 \cdot \cos\left(\frac{2\pi f \Delta l}{c} + \varphi_0  \right) \right]
\end{align}
where $p_{1,2}$ are the optical powers at each photodetector in mid-fringe, $c_{1,2}$ are the interferometric contrasts at each photodetector, $\Delta l$ is the interferometer's optical path length difference, $c$ is the speed of light, and $\varphi_0$ is an arbitrary constant. After the TIA, the resulting signal is given by
\begin{align}
v(f) &= G \left[ P_1(f) - P_2(f) \right] \nonumber\\
&= G \left[ p_1 - p_2  + (c_1 p_1 + c_2 p_2) \cdot \cos \left( \frac{2\pi f \Delta l}{c} + \varphi_0 \right)\right]
\end{align}
where $G\,[\mathrm{V}/\mathrm{W}]$ is the trans-impedance gain. In order to attain balanced operation (i.e., $p_1 = p_2 $), we leverage the reflectivity dependence of BS4 on the macroscopic beam incidence angle, achieving nearly equal power levels on both photodiodes, such that
\begin{equation}
v(f) = G p_1 (c_1 + c_2 ) \cdot \cos \left( \frac{2\pi f \Delta l}{c} + \varphi_0 \right)
\label{equation:mzi}
\end{equation} 
Equation~\ref{equation:mzi} has periodic zero crossings that we use for laser locking. The slope of the error signal at the operating point is proportional to the available optical power, the interferometric contrasts, the trans-impedance gain, and the interferometer's arm length difference.

A Moku:Lab by Liquid Instruments~\cite{moku} is used as a digital controller to provide feedback based on the generated error signal to both the slow thermal actuator and the fast piezo-electric transducer actuator of the NPRO laser. Additionally, a pre-amplifier (SR560 by Stanford Research Systems) and a post-amplifier equipped with a low-pass filter are used to mitigate analog-to-digital converter noise originating from the digital servo and to enhance the low-frequency gain, respectively.

When the laser is locked to the MZI, the optical pathlength stability of the interferometer is transferred to the frequency stability of the laser, obeying Equation~\ref{equation:dl-to-df}. To isolate the MZI from external perturbations affecting its pathlength noise $\delta l$, it is placed inside a vacuum chamber at a moderate pressure of $10^{-6}\,$mbar and surrounded by a set of three aluminum heat shields (Figure~\ref{figure:MZI}b), similar to the systems designed for high-performance metrology with ultra-stable optical reference cavities~\cite{Dovale2019PhD, Sanjuan2015}. 

Each shield consists of six 10\,mm aluminum plates fastened together via M4 screws, and is supported by three 10\,mm PEEK spheres that rest on 5\,mm-deep conical cutouts made to each of the baseplates, with the exception of the outermost shield, that rests on three PEEK semi-spheres placed on the surface of the vacuum chamber. The plates are polished to yield a low emissivity, thereby slowing radiative heat transfer inside the enclosure. The resulting system has an extremely slow response to temperature changes, with a thermal time constant of about one week that is largely dominated by a surface-to-surface radiative exchange.

Laser A is split two ways. One part of the light feeds the vacuum chamber, where it is further split such that a small fraction is captured by a photoreceiver for amplitude stabilization of the laser and the rest is injected into the MZI. The other part is split two ways and interfered with two reference lasers to enable measurements of the achieved stability. The reference systems, lasers B and C, are two iodine-stabilized NPRO lasers (Prometheus by Coherent), locked to the molecular iodine hyperfine transitions R(56)32-0 `a1' and `a2'. The two reference lasers are also interfered, generating a third beatnote signal that allows us to perform a complete characterization of the three systems' stability. The three beatnotes are captured by high-speed InGaAs photoreceivers. The two beatnotes with laser A, which are in the 0.5-2\,GHz band, are mixed down to less than 100\,MHz using an ultra-stable GHz signal generator (SMB100A by Rohde \& Schwarz). Finally, the three beatnotes are tracked simultaneously by a Moku:Pro phasemeter~\cite{moku}. The noise contributions of the R\&S SMB100A and the phasemeter instrument are measured to be well below 1\,Hz$/\!\sqrt{\mathrm{Hz}}$ throughout the whole band. 

\section{Results}

\begin{figure}[t!]
\centering
\xincludegraphics[width=0.45\textwidth, label=a), fontsize=\large]{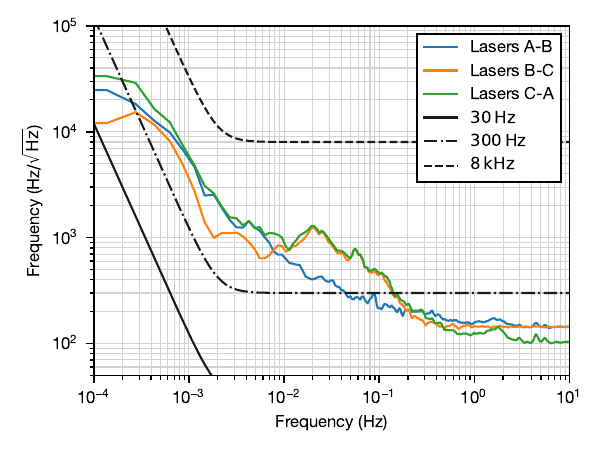}
\xincludegraphics[width=0.45\textwidth, label=b), fontsize=\large]{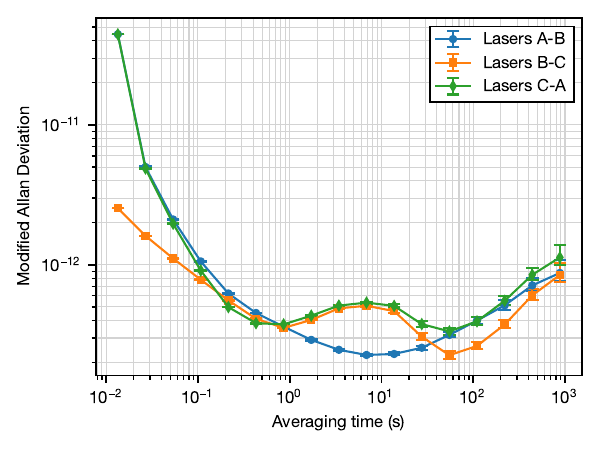}
\caption{Frequency spectral densities (a) and modified Allan deviations (b) of a 2-hour measurement of the three beatnotes (A: MZI-stabilized laser; B, C: iodine-stabilized lasers). Also shown are the frequency noise spectral densities of 30\,Hz$/\!\sqrt{\mathrm{Hz}} \cdot u(f)$, 300\,Hz$/\!\sqrt{\mathrm{Hz}}\cdot u(f)$, and 8\,kHz$/\!\sqrt{\mathrm{Hz}}\cdot u(f)$.}
\label{figure:3beats}
\end{figure}

The frequency spectral densities~\cite{Troebs2006} and modified Allan deviations~\cite{NIST1065} of the three beatnotes are shown in Figure~\ref{figure:3beats}a and~\ref{figure:3beats}b respectively for a typical 2-hour measurement at a rate of 150 samples per second. Also shown in Figure~\ref{figure:3beats}a is the frequency noise spectral density of 30\,Hz$/\!\sqrt{\mathrm{Hz}}$, 300\,Hz$/\!\sqrt{\mathrm{Hz}}$, and 8\,kHz$/\!\sqrt{\mathrm{Hz}}$, scaled by Equation~\ref{equation:noise-shape}, respectively representing the stability of a space-qualified cavity-stabilized laser system, the LISA laser frequency pre-stabilization target, and the picometer-equivalent frequency instability of an interferometer with 7\,cm arm length difference (see Equation~\ref{equation:dl-to-df}).

Inspection of any of these two plots, containing largely the same information, reveals that the MZI-stabilized system (laser A) offers a stability comparable to the two iodine-stabilized systems (lasers B, C). This complicates the assessment of the achieved stability since the frequency noise of the reference lasers cannot be neglected in comparison to the frequency noise of the unit under test.

A bump in the noise of the B-C and the C-A beatnotes is evident in both the amplitude spectral density estimates and the modified Allan deviations. The bump can be seen at frequencies between $100\,$mHz and 0.2\,Hz, or at averaging times between $2$ and $50$ seconds. From this analysis, it is not clear which of lasers A or C is responsible for the noise degradation.

To evaluate the performance of each single laser, independently from the rest, we carry out a three-cornered-hat analysis~\cite{Gray1974} using the modified Allan deviation (MDEV) and the Hadamard deviation (HDEV) as the statistic functions of choice~\cite{NIST1065}. 

The MDEV is chosen for its ability to distinguish between white and flicker phase noise at short averaging times (i.e., at short $\tau = m \tau_0$, where $\tau$ is the averaging time, $\tau_0$ is the gate time or sampling time, and $m$ is the averaging factor), or equivalently at high frequencies. The MDEV is also widely used in the time and frequency standards community, such that our stability results may be easily compared to other references.

The HDEV is chosen for its ability to handle divergent noise sources at long averaging times. The MDEV is not a good statistic for processes having power spectral densities with $f^{-4}$ dependency (e.g., as Equation~\ref{equation:noise-shape}), as the obtained variance at long $\tau$ may depend on the measurement time. On the other hand, the HDEV examines the second difference of the fractional frequencies, which makes it robust against $f^{-4}$ noise, allowing a direct comparison of the achieved long-term stability with noise allocations following Equation~\ref{equation:noise-shape}.

\begin{figure}[t!]
\centering
\includegraphics[width=9cm]{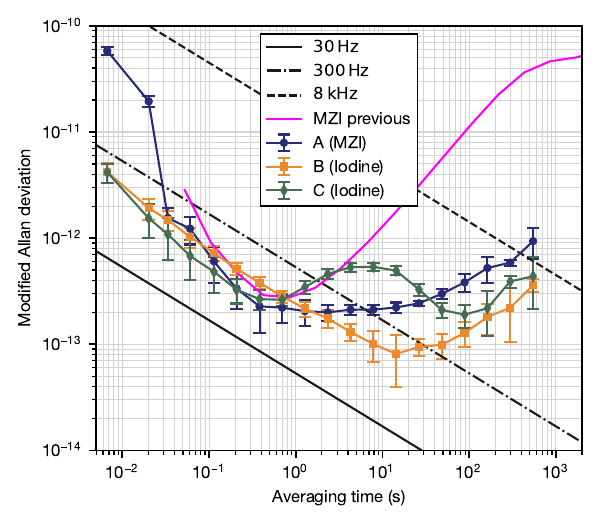}
\caption{Modified Allan deviation of the three stabilized lasers derived from a three-cornered hat analysis. The data show the average instability of 10 data sets with a duration of 1.2 hours each. Error bars represent the standard deviation of the data averaged for each point. Also shown are the modified Allan deviations of virtual beatnotes with white frequency noise at 30\,Hz$/\!\sqrt{\mathrm{Hz}}$, 300\,Hz$/\!\sqrt{\mathrm{Hz}}$, and 8\,kHz$/\!\sqrt{\mathrm{Hz}}$.}
\label{figure:tch-mdev}
\end{figure}

\begin{figure}[t!]
   \centering
   	\includegraphics[width=9cm]{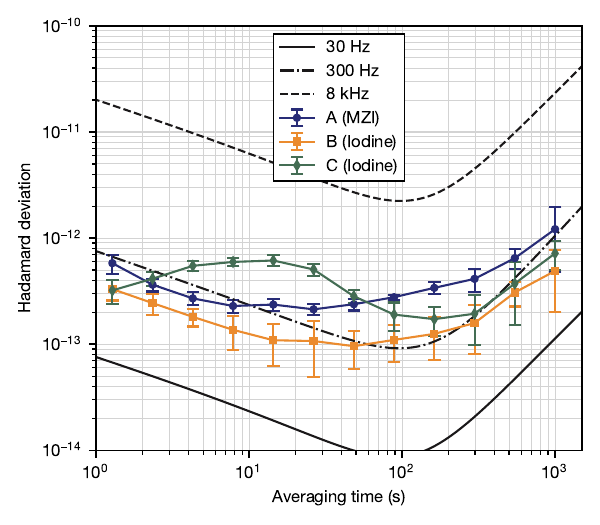}
   		\caption{Hadamard deviation of the three stabilized lasers derived from a three-cornered hat analysis. The data show the average instability of 10 data sets with a duration of 1.2 hours each. Error bars represent the standard deviation of the data averaged for each point. Also shown are the Hadamard deviations corresponding to virtual beatnotes having frequency spectral densities of 30\,Hz$/\!\sqrt{\mathrm{Hz}} \cdot u(f)$, 300\,Hz$/\!\sqrt{\mathrm{Hz}}\cdot u(f)$, and 8\,kHz$/\!\sqrt{\mathrm{Hz}}\cdot u(f)$.}
   		\label{figure:tch-hdev}
\end{figure}

\begin{figure}[t!]
\centering
\includegraphics[width=9cm]{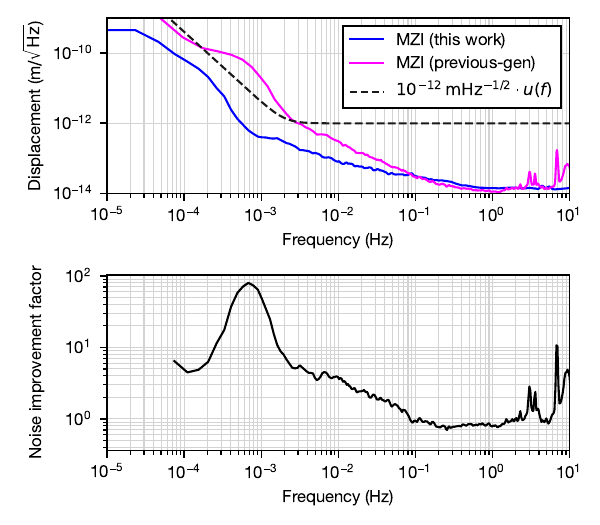}
\caption{Spectral density of optical path length noise in the MZI from a 12-hour measurement, as probed by laser B. The noise floor is limited at the lower frequencies by temperature fluctuations coupling as interferometer path length changes. At frequencies higher than 1\,mHz, a $1/f$ power spectrum is observed, which is suspected to be due to the effect of thermal drifts in the sensitive trans-impedance amplifier used to convert the differential interferometric current into a voltage for laser locking. Also shown are the previously realized stability, and the $1\,\mathrm{pm}/\!\sqrt{\mathrm{Hz}} \cdot u(f)$ displacement noise allocation commonly used for the local interferometry in LISA.}
\label{figure:displacement}
\end{figure}

Frequency data were taken over 12\,h by the same phasemeter with a gate time of 6.7\,ms. The data set was cut into 10 sections of 1.2\,h, a linear drift was removed for compensating the temperature drift in each section, and the individual Allan deviations were calculated. The arithmetic mean and standard deviation of the resulting modified Allan deviation are shown in Figure~\ref{figure:tch-mdev}, which also shows the MDEV of the MZI stability reported in~\cite{Gerberding2017}, obtained via a domain conversion from the power spectral density using the MDEV transfer function~\cite{Rubiola2008}, and the MDEV of virtual beatnotes having white frequency noise spectral densities at the 30\,Hz$/\!\sqrt{\mathrm{Hz}}$, 300\,Hz$/\!\sqrt{\mathrm{Hz}}$, and 8\,kHz$/\!\sqrt{\mathrm{Hz}}$ levels. 

Finally, given that we are only interested in the Hadamard deviation at long averaging times, the data is downsampled by a factor of 100 in order to decrease the otherwise long computation time of the HDEV. To downsample the data, an averaging operation is performed, which limits the results to $\tau > 100 \cdot \tau_0$. The results are shown in Figure~\ref{figure:tch-hdev}, together with the HDEV corresponding to virtual beatnotes with 30\,Hz$/\!\sqrt{\mathrm{Hz}} \cdot u(f)$, 300\,Hz$/\!\sqrt{\mathrm{Hz}}\cdot u(f)$, and 8\,kHz$/\!\sqrt{\mathrm{Hz}}\cdot u(f)$ frequency spectral densities, obtained by numerically computing the HDEV on data generated by a simple noise model that conforms to Equation~\ref{equation:noise-shape}. 

The three-cornered-hat analysis reveals the individual performance of each laser. The noise bump of laser C is successfully isolated at $\tau$ between 1 and 10\,s. At $\tau > 10\,$s, laser B is roughly a factor 2 to 3 more stable than laser A. At lower averaging times, their stability is very similar, except at very short averaging times ($\tau \sim10$\,ms), where the MZI suffers from short-lived instabilities originating from the coupling of vibrations of the vacuum pumps.

As per the obtained MDEV, the MZI's fractional frequency instability is below $10^{-12}$ at averaging times greater than 0.1\,s and in excess of a few hundred seconds. A maximal stability of $2 \times 10^{-13}$ is achieved between 1 and 10\,s that is dominated by flicker frequency noise (i.e., $1/f$ noise). In this range, the MZI is compatible with the LISA frequency pre-stabilization target (Equation~\ref{equation:freq-noise}). On the other hand, the HDEV analysis reveals that the MZI does not meet the LISA target at averaging times between 10 and a few hundred seconds (e.g., it is a factor of 2 less stable at 50 seconds). At 1000 seconds, the MZI stability is close to the target, hinting that at even longer measurement times, the system may be compliant with this noise allocation, which can only be revealed by performing longer measurements.

Using Equation~\ref{equation:dl-to-df} we can infer the achieved MZI pathlength stability as probed by laser B, shown in Figure~\ref{figure:displacement}, and compare it against the previous results~\cite{Gerberding2017}. Thanks to the three-cornered-hat analysis we have confidence in the estimated stability at the lower frequencies, e.g., at frequencies below 100\,mHz, where it was shown that laser B is, on average, 2.4 times more stable than laser A. At higher frequencies (e.g., around 1\,Hz), the stabilities of the three lasers are similar, so the performance of laser A may be estimated as $\frac{1}{2}$ of the stability of the A-B or C-A beatnotes, which yields a noise floor of 7\,fm$/\!\sqrt{\mathrm{Hz}}$ at 1\,Hz. These results showcase a clear performance improvement of the setup below 100\,mHz.	

\section{Conclusion}

An ultra-stable Mach-Zehnder interferometer with unequal arm lengths capable of reaching $1\,\mathrm{pm}/\!\sqrt{\mathrm{Hz}} \cdot u(f)$ from $10\,\upmu$Hz to 10\,Hz was presented. The new system is one to two orders of magnitude more stable in the lower frequencies than the previous realization. This stability was achieved by applying a combination of two passive techniques for reducing noise sources of thermal origin. First, a quasi-monolithic fiber injector made of fused silica components provides an ultra-stable input beam that is much more robust to temperature changes than what is possible with conventional fiber injectors. Second, a high-performance enclosure provides an ultra-quiet thermal environment, reducing the coupling of temperature-driven effects to path length noise in the interferometer, such as thermoelastic deformation of the baseplate and components, and refractive index fluctuations. 

A combination of frequency and time domain analysis techniques was used to assess the stability of the MZI-stabilized laser along with two iodine-stabilized reference lasers. The individual stability of each system was disentangled from a simultaneous three-signal measurement using the three-cornered-hat method. Due to the nature of the involved noise sources, which are white frequency noise at high frequency (i.e., at a short averaging time, $\tau$), and random run noise at low frequency (i.e., at long $\tau$), two different variance functions were used, with one providing greater confidence at short $\tau$ (the modified Allan deviation), and one greater confidence at long $\tau$ (the Hadamard deviation). 

The three-cornered-hat analyses revealed that the stability of the MZI system is comparable to the two reference systems, which are based on stabilization to molecular iodine hyperfine transitions near 532\,nm. To the best of the authors' knowledge, the achieved stability constitutes a record for compact laser interferometers~\cite{Watchi2018}.

The frequency of our MZI-stabilized laser is within the target of $300\,\mathrm{Hz}/\!\sqrt{\mathrm{Hz}} \cdot u(f)$ for all frequencies above 40\,mHz. At frequencies between 10\,$\upmu$Hz and 40\,mHz, it is a factor of 1 to 5 less stable. The stability could be improved by increasing the interferometer's arm length difference (e.g., from $7$\,cm to $40$\,cm, as in the LTP interferometer), and addressing the associated complexities of the longer optical path length. However, a more exciting prospect is combining the techniques of Mach-Zehnder stabilization and arm-locking~\cite{LFCST}, which could lead to a frequency stability orders of magnitude better than $300\,\mathrm{Hz}/\!\sqrt{\mathrm{Hz}} \cdot u(f)$, potentially allowing the requirements on time delay interferometry to be relaxed. Since the reference interferometer can be integrated as part of the optical bench that is already a central feature in this type of mission, this technique eliminates the need for a separate laser stabilization subsystem, which makes this an interesting scheme for future gravity missions~\cite{Luo2016, Luo2020}.

%

\section*{Acknowledgements}

The authors would like to thank Germ\'an Fern\'andez Barranco for his help with analog electronics, and Oliver Gerberding and Katharina-Sophie Isleif for their continued cooperation in the project. M.D.A.\ would like to thank Olaf Hartwig for fruitful discussions on time-domain stability analysis.

This work was funded by the Deutsche Forschungsgemeinschaft (DFG, German Research Foundation) Project-ID 434617780-SFB 1464. The authors acknowledge support from the Deutsche Forschungsgemeinschaft (DFG) Sonderforschungsbereich 1128 Relativistic Geodesy and Cluster of Excellence "QuantumFrontiers: Light and Matter at the Quantum Frontier: Foundations and Applications in Metrology" (EXC-2123, Project No.\ 390837967) and Max Planck Society (MPS) through the LEGACY cooperation on low-frequency gravitational wave astronomy (M.IF.A.QOP18098). 

The authors also acknowledge support by the German Aerospace Center (DLR) with funds from the Federal Ministry of Economics and Technology (BMWi) according to a decision of the German Federal Parliament (Grant No.\ 50OQ2301, based on Grants No.\ 50OQ0601, No.\ 50OQ1301, No.\ 50OQ1801).

\end{document}